\documentclass[conference,letter]{IEEEtran}
\IEEEoverridecommandlockouts
\usepackage{cite}
\usepackage{amsmath,amssymb,amsfonts}
\usepackage{algorithm}
\usepackage{algorithmic}
\usepackage{graphicx}
\usepackage{textcomp}
\usepackage{xcolor}
\usepackage{bbm}
\usepackage{booktabs}
\usepackage{subfig}

\def\BibTeX{{\rm B\kern-.05em{\sc i\kern-.025em b}\kern-.08em
    T\kern-.1667em\lower.7ex\hbox{E}\kern-.125emX}}
\usepackage{glossaries}
\usepackage{caption}
\newacronym{iot}{IoT}{Internet of Things}
\newacronym{hmm}{HMM}{Hidden Markov Model}
\newacronym{bsc}{BSC}{Binary Symmetric Channel}
\newacronym{ldpc}{LDPC}{Low Density Parity Check}
\newacronym{bch}{BCH}{Bose-Chaudhuri-Hocquenghem}
\newacronym{ira}{IRA}{Irregular Repeat Accumulate}
\newacronym{map}{MAP}{Maximum A Posteriori}
\newacronym{harq}{HARQ}{Hybrid Automatic Repeat Request}
\title{Hidden Markov Model-Based Encoding for Time-Correlated IoT Sources}

\author{\IEEEauthorblockN{Siddharth Chandak}
\IEEEauthorblockA{\textit{Department of Electrical Engineering} \\
\textit{IIT Bombay}, Mumbai, India \\
email: chandak1299@iitb.ac.in}
\and
\IEEEauthorblockN{Federico Chiariotti}
\IEEEauthorblockA{\textit{Department of Electronic Systems} \\
\textit{Aalborg University}, Aalborg, Denmark \\
email: fchi@es.aau.dk}
\and
\IEEEauthorblockN{Petar Popovski}
\IEEEauthorblockA{\textit{Department of Electronic Systems} \\
\textit{Aalborg University}, Aalborg, Denmark \\
email: petarp@es.aau.dk}
}

\begin{document}

\maketitle

\begin{abstract}
As the use of \gls{iot} devices for monitoring purposes becomes ubiquitous, the efficiency of sensor communication is a major issue for the modern Internet. Channel coding is less efficient for extremely short packets, and traditional techniques that rely on source compression require extensive signaling or pre-existing knowledge of the source dynamics. In this work, we propose an encoding and decoding scheme that learns source dynamics online using a \gls{hmm}, puncturing a short packet code to outperform existing compression-based approaches. Our approach shows significant performance improvements for sources that are highly correlated in time, with  no additional complexity on the sender side.
\end{abstract}


\glsresetall

\section{Introduction}

Over the last few years, the vision of the \gls{iot} has become a reality: all over the world, billions of wireless sensors gather and transmit data about all facets of our lives~\cite{forecast2019cisco}, from Smart City applications such as live road traffic and parking spot mapping to industrial ones such as the structural monitoring of buildings and bridges, or even environmental ones such as weather and pollution measurement. In this context, the traditional Internet packet structure can become inefficient: the short length of the packet payload makes the overhead from protocol headers become more significant~\cite{shi2020low}, and channel codes need to be designed for the short packet regime~\cite{wu2020pilot}. 

However, \gls{iot} sensors have some common features that can be exploited to increase transmission efficiency: as the processes they monitor are often slow-varying and highly correlated in time~\cite{rastogi2020iot,shirbhate2018time}, they can be represented with \glspl{hmm}~\cite{rabiner1989tutorial}. The overall objective of this work is to show how learnable features of the source can decrease the communication requirements as learning progresses and a Markov source is one of the simplest ways to show this effect. The standard approach to exploit some regularities in the source data is is source compression~\cite{stojkoska2017data}: as packets are highly correlated in time, the sensed data can be compressed using previously transmitted data as a reference, reducing the amount of data sent with no loss of accuracy~\cite{biason2017ec}. However, compression-based schemes have two drawbacks: firstly, they need to be shared by both endpoints of the communication. If the source's statistics are time-varying, and the compression needs to be adapted to its changes over time, the transmitter and receiver need to exchange a significant amount of signaling for the latter to be able to decode messages. Secondly, the estimation of the source statistics needs to be performed by the source itself, i.e., the battery-powered \gls{iot} sensor. This operation can be computationally expensive, reducing or even canceling the energy efficiency gains in the communication.

The decoder can also be adapted to exploit information from previous packets, obtaining the \gls{map} decoding: the optimal \gls{ldpc} decoder for Markovian sources was derived in~\cite{islam2018information}. In the case of variable length encoding~\cite{thobaben2005low}, the source-aware decoder becomes more efficient in the case of long symbol sequences, as its does not increase the size of the decoding trellis. If the Markov model is not known at the beginning, the authors of \cite{majumder2012joint} propose an implementation of \gls{ira} codes that can take into account its learned transition probability in the message passing process. In this work, we take a similar approach: by modeling the temporally correlated source as an \gls{hmm}, we can gradually learn the statistics of the underlying process at the receiver side, allowing a \gls{map} decoding of messages sent over a noisy channel. Instead of compressing messages to remove redundant information (i.e., the correlation between subsequent packets), we exploit it at the receiver side to improve decoding, and puncture short \gls{bch} codes~\cite{bose1960class} to increase efficiency.  \gls{bch} is a practical explicit coding scheme, which can give better results with compression schemes than the short packet coding bounds, which are not associated with a practical code construction.  Our proposed modifications in encoding and decoding are independent of a specific scheme and will work well with any other code.

In our scheme, we divide messages in a correlated state and an uncorrelated message, then remove the state bits from the encoded message by puncturing. The punctured part of the message can be recovered by using the parity bits and the side information from previous packets. Previous works had explored \gls{map} decoding for temporally correlated sources for turbo codes~\cite{zhao2006turbo} and \gls{ldpc} codes~\cite{garcia2003ldpc}, but to the best of our knowledge this is the first work to propose a scheme that exploits time correlation in the source for short packets.  We assume no initial knowledge of the Markov transition probabilities, which are learned on the fly by the transmitter and receiver.  Our scheme significantly outperforms traditional encoding and compression schemes as long as the source shows significant correlation, both for ergodic and time-varying sources whose transition probabilities change over time. Furthermore, unlike other \gls{hmm} schemes for longer codes, it requires no additional complexity on the sender side, relieving battery-powered \gls{iot} node of the additional computational complexity, and the consequent energy expenditure, that efficient source compression requires. 

The rest of this paper is divided as follows: Sec.~\ref{sec:system} presents the model of the encoder and decoder and the proposed coding scheme, while Sec.~\ref{sec:results} shows the simulation results for all considered techniques. Finally, Sec.~\ref{sec:concl} concludes the paper and lists some possible avenues for future work. 

\section{System Model}\label{sec:system}
Let the information source be represented by a Markov chain with state space $\mathcal{S}$, whose cardinality is $S$. The Markov chain is defined by its $S\times S$ sparse transition matrix $T$. When the source is in state $s$, it randomly generates messages from set $\mathcal{M}(s)$; we assume that the message set size is the same for all states, i.e., $|\mathcal{M}(s)|=M, \forall s\in\mathcal{S}$. Each message $m\in\mathcal{M}(s)$ is generated with uniform probability $\frac{1}{M}$. The entropy of the message is then $\log_2(M)$ bits, and the source state entropy is bounded by $\log_2(S)$. Hence, the number of information bits required to represent a message is $k=\log_2(S)+\log_2(M)$. The information source needs to send the message to a receiver over a \gls{bsc} with bit error probability $p_b$, so it uses a \gls{bch} code: codewords are composed of $n\geq k$ bits. 

\begin{figure}[t]
\begin{center}
\vspace{0.05in}
  \includegraphics[width=0.8\columnwidth]{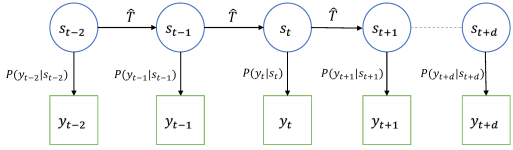}
  \caption{Hidden Markov model for decoding: $S$ hidden states, $2^n$ observed states (received packets) and transition matrix $\widehat{T}$}
  \label{Fig:HMM_fig}
\end{center}
\end{figure}

In traditional networking, each packet is encoded independently, and the encoding assumes that the state $s$ has a uniform distribution. However, this does not exploit the temporal correlation in the system state. It is possible for the receiver to use the transition matrix of the source as additional information to improve the decoding process. Let the codeword obtained on encoding state $s\in\mathcal{S}$ and $m\in\mathcal{M}(s)$ be $x(s,m)$ and $\mathcal{X}(s)$ be the set of codewords corresponding to state $s$.
We can then see this as an \gls{hmm}, as shown in Fig.~\ref{Fig:HMM_fig}: the received packet at time $t$ is denoted as $y_t\in\mathcal{Y}$, can be seen as an observation of the hidden state, which corresponds to the state of the source $s_t$. We can then get the emission probabilities for the \gls{hmm} as follows:
\begin{equation}
        P(Y_t=y_t|S_t=s_t)=\frac{1}{M}\underset{x\in\mathcal{X}(s_t)}\sum\left(\frac{p_b}{1-p_b}\right)^{d_{xy}}(1-p_b)^n,
    \end{equation}
where $d_{xy}$ is the Hamming distance between $x$ and $y$. 
We use Forward algorithm \cite{rabiner1989tutorial} for \gls{map} decoding of the packets. Formally, we can define this by obtaining prior and posterior probabilities. The probability of getting a received word $y$ when $x$ is sent is given by:
\begin{equation}
    p_{Y|X}(y|x)=\left(\frac{p_b}{1-p_b}\right)^{d_{xy}}(1-p_b)^n.
\end{equation}
If we assume that the receiver has a prior distribution $\hat{P_t}$ for the state of the system, the a posteriori decoding probability is:
\begin{equation}
\begin{aligned}
    p'_{X_t|Y_t,\hat{P_t}}(x|y_t,\hat{p}_t)&=\frac{p_{X_t|\hat{P}_t}(x|\hat{p}_t)p_{Y_t|X_t}(y_t|x)}{p_{Y_t|\hat{P}_t}(y_t|\hat{p}_t)}\\
    &=\frac{\hat{p}_t(s(x))\left(\frac{p_b}{1-p_b}\right)^{d_{xy_{_t}}}}{\underset{s\in\mathcal{S}}\sum \underset{x'\in\mathcal{X}(s)}\sum\hat{p}_t(s) \left(\frac{p_b}{1-p_b}\right)^{d_{x'y_{_t}}}},
\end{aligned}
\end{equation}
where $s(x)$ is the state to which $x$ belongs. For states, the a posteriori probability can be calculated as:
\begin{equation}
\begin{aligned}
p'_{S_t|Y_t,\hat{P}_t}(s|y_t,\hat{p}_t)&=\frac{\hat{p}_t(s)p_{Y_t|S_t}(y_t|s)}{p_{Y|\hat{P}_t}(y_t|\hat{p}_t)}\\
&=\frac{\hat{p}_{t}(s)\underset{x\in\mathcal{X}(s)}\sum\left(\frac{p_b}{1-p_b}\right)^{d_{xy_{_t}}}}{\underset{s'\in\mathcal{S}}\sum\underset{x'\in\mathcal{X}(s)}\sum\hat{p}_t(s') \left(\frac{p_b}{1-p_b}\right)^{d_{x'y_{_t}}}}.
\end{aligned}
\end{equation}
We can define the \emph{a priori} probability as:
    \begin{equation}
        \hat{p}_{t+1}(s)=\underset{{s'\in\mathcal{S}}}\sum(T_{s's}p'_{S_t|Y_t,\hat{P}_t}({s'}|y_t,\hat{p}_t)).
    \end{equation}
The state and message are then inferred as follows:
\begin{equation}
    \hat{s_t}=\underset{s\in\mathcal{S}}{\arg\max} \left(p'_{S_t|Y_t,\hat{P}_t}(s|y_t,\hat{p}_t)\right);
\end{equation}
\begin{equation}
    \hat{m_t}=\underset{m\in\mathcal{M}(\hat{s_t})}{\arg\min} (d_{x(\hat{s_t},m)y_{_t}}).
\end{equation}

\subsection{Delayed MAP decoding}

We can also consider delayed \gls{map} decoding, including both previous packets and future ones. Naturally, considering future packets requires allowing a delay of $d$ packets to find the \gls{map} probability $P(s_t|y_1,y_2,...,y_t,...,y_{t+d})$ using a forward-backward algorithm \cite{rabiner1989tutorial} on the Markov model. For a given delay $d$, the \emph{a priori} probability for any sequence of states $\mathbf{s}=\{s_t,s_{t+1},...,s_{t+d}\} \in \mathcal{S}^{d+1}$ with prior $\hat{p}_t(s)$ (from above) can be calculated as:
\begin{equation}
    \hat{p}(\mathbf{s})=\hat{p}_t(s)\prod_{i=t+1}^{t+d}T_{s_{i-1}s_i}
\end{equation}
The conditional probability of the received sequence $\mathbf{y}=\{y_t,y_{t+1},...,y_{t+d}\}$ is:
\begin{equation}
    p_{\mathbf{Y}|\mathbf{S}}(\mathbf{y}|\mathbf{s})=\prod_{i=t}^{t+d}\left(\frac{1}{M}\sum_{x\in\mathcal{X}(s_i)}\left(\frac{p_b}{1-p_b}\right)^{d_{xy_{_i}}}\right)
\end{equation}
The a posteriori probability of the sequence  and statecan be obtained using Bayes' theorem:
\begin{align}
    p_{\mathbf{S}|\mathbf{Y},\hat{P}}(\mathbf{s}|\mathbf{y},\hat{p})&=\frac{p_{\mathbf{Y}|\mathbf{S}}(\mathbf{y}|\mathbf{s})\hat{p}(\mathbf{s})}{\underset{{\mathbf{s}'\in\mathcal{S}^{d+1}}}\sum\hat{p}(\mathbf{s}')p_{\mathbf{Y}|\mathbf{S}}(\mathbf{y}|\mathbf{s'})}\\
    p_{S_t|\mathbf{Y},\hat{P}}(s_t=s|\mathbf{y},\hat{p})&=\sum_{\mathbf{s}\in\mathcal{S}^{d+1}:s_t=s}p_{\mathbf{S}|\mathbf{Y},\hat{P}}(\mathbf{s}|\mathbf{y},\hat{p})
\end{align}
We can then infer the most likely state and message as above, using the additional information. If the decoding at time $t$ is extremely uncertain, the next packet might be enough to remove the uncertainty, using the information from the next packet to infer \emph{a posteriori} the state of the system at time $t$.  The \gls{map} decoding approach is not commonly used in decoding because of its computational cost, as the decoder needs to evaluate all possible codewords at every decoding, but it is practical for short-packet codes with a limited number of options, with $O(MS^2)$ complexity.

\subsection{Coding schemes}

The \gls{map} approach can reduce the packet error rate by exploiting the time-dependent information, particularly so when using a forward-backward solution with delayed decoding, but it is possible to improve performance even further by proper code design by the sender. 

Traditionally, this can be accomplished with source compression, encoding states into fewer than $\log_2(S)$ bits based on the Markov chain transition matrix. Considering the same packet size $n$, source compression of the state can increase protection by using more parity bits. We considered three different legacy schemes based on source compression:
\begin{enumerate}
    \item \textbf{Legacy encoding}: This scheme does not use any temporal information on the source in the encoding. The $k$-bit payload representing the state and message are concatenated and then encoded using \gls{bch} into $n$ bits.
    \item \textbf{Stationary compression encoding}: The state bits are compressed using Huffman coding based on the stationary distribution $\Pi$ of the Markov chain with transition matrix $T$. We obtain a variable length code for states. These bits are concatenated with message bits and encoded using \gls{bch}, which is then punctured to obtain a codeword of length $n$. The decoding schemes mentioned above can be directly applied here.
    \item \textbf{Conditional compression encoding}: Suppose we are in state $s_{t-1}$ at time $t-1$. The state at time $t$ is compressed using Huffman coding based on the conditional distribution $T_{s_{t-1}s}=P(S_t=s|S_{t-1}=s_{t-1})$. Since the transition matrix is assumed to be sparse, using the conditional distribution for coding requires fewer bits. Every $t_c$-th packet is a check packet which is a packet of the same size but is encoded using combined encoding (the state bits are not compressed).
\end{enumerate}

\begin{figure}[t!]
\begin{center}
  \includegraphics[width=0.7\columnwidth]{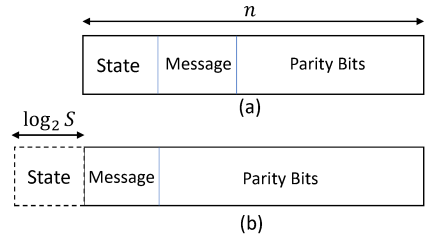}
  \caption{(a) Legacy encoding. (b) Punctured encoding (dashed box represents punctured bits)}
  \label{Fig:Encoding_fig}
\end{center}
\end{figure}

Our \textbf{punctured encoding} scheme does not act on source compression, but rather changes the coding scheme to provide additional protection to the message. The state and message are concatenated to form a $k$-bit payload and then encoded using \gls{bch} into $(n+\log_2(S))$ bits. The $\log_2(S)$ state bits are then removed to get a codeword of $n$ bits. By puncturing the state bits, we provide more protection to the message bits.
The proposed scheme, along with the legacy coding, is shown in Fig.~\ref{Fig:Encoding_fig}.

\subsection{Learning Parameters}
 Since we do not assume any prior knowledge of the scenario, the receiver needs to learn the \gls{bsc} parameter $(p_b)$ and the transition matrix for the Markov source. The bit flip probability for the channel can be estimated by beginning the transmission of updates using a highly redundant encoding. 
\begin{equation}
    \hat{p_b}=\frac{\text{Number of bit flips}}{\text{Total bits sent}}.
\end{equation}
In the beginning of communication, the receiver assumes a uniform transition matrix. Since the statistics of the source are unknown to the receiver, the sender cannot use compression encoding, but legacy or punctured encoding work, although with lower performance. Each received packet is decoded using either of the above mentioned decoding schemes. The estimated transition matrix is then updated based on the observed transitions.
\begin{equation}
    \widehat{T}_{s_is_j}=\frac{N_{s_is_j}+\alpha}{\underset{s'\in\mathcal{S}}{\sum}(N_{s_is'}+\alpha)}.
\end{equation}
where $\alpha$ is a non-zero parameter which can be varied to vary the speed of learning of parameters and $N_{s_is_j}$ is the number of transitions observed from state $s_i$ to state $s_j$. The transition matrix and channel error probability is updated online by using the estimators over a sliding window of 1000 packets, in order to deal with fluctuations in the source and channel.

\section{Results}\label{sec:results}
We simulated the communication strategies from the previous section using a Monte Carlo approach. We obtained the steady state performance by simulating 10 sequences of $10^5$ packets each. We considered a packet size of $n=20$ bits, with $S=32$ and $M=32$.
The transition matrix for the Markov chain is generated randomly for each sequence as a sparse matrix. We study the variation in packet errors by varying the bit error probability $p_b$, the density of the transition matrix and the decoding delay. The density of the sparse matrix is defined as the number of non-zero elements over the total number of elements. It is directly proportional to the entropy rate of the Markov source. The results showing the parameter learning transient were averaged over $500$ sequences. 

\begin{figure}[t!]
\begin{center}
  \includegraphics[width=0.9\columnwidth]{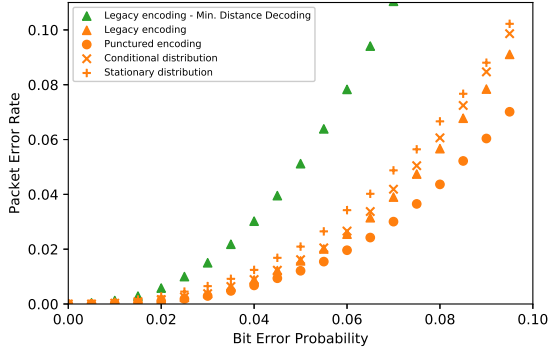}
  \caption{Packet error rate as a function of $p_b$ for various encoding schemes with MAP decoding ($T$ is a sparse matrix with density 0.125)}
  \label{Fig:MAP}
 \end{center}
\end{figure}

\begin{figure}[t!]
\begin{center}
  \includegraphics[width=0.9\columnwidth]{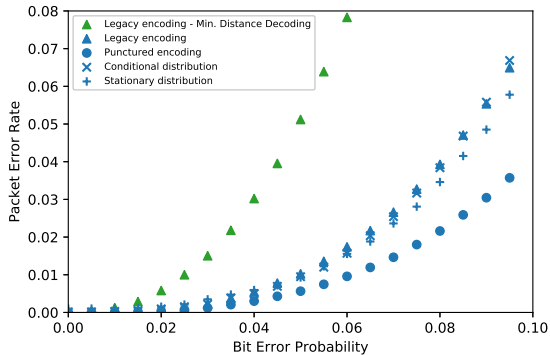}
  \caption{Packet error rate as a function of $p_b$ for various encoding schemes with Delayed decoding (delay $d=1$ and $T$ is a sparse matrix with density 0.125)}
  \label{Fig:Delayed}
  \end{center}
\end{figure}

We first consider the steady-state performance of the schemes, evaluated in terms of packet error probability as a function of the \gls{bsc} bit error probability $p_b$. Fig.~\ref{Fig:MAP} and Fig.~\ref{Fig:Delayed} show performance of different encoding schemes for \gls{map} and delayed decoding respectively. As expected, packet errors increase with higher bit error probability and both \gls{map} and delayed decoding significantly outperform minimum distance decoding. Punctured encoding has a significant advantage over the traditional encoding schemes when we use delayed decoding, and a smaller but still significant advantage in case of \gls{map} decoding, as it can make better use of the temporal correlation between successive states. The results for conditional compression encoding are for $t_c=2$, as it was the setting that gave the best performance.

\begin{figure}[t!]
\begin{center}
  \includegraphics[width=0.9\columnwidth]{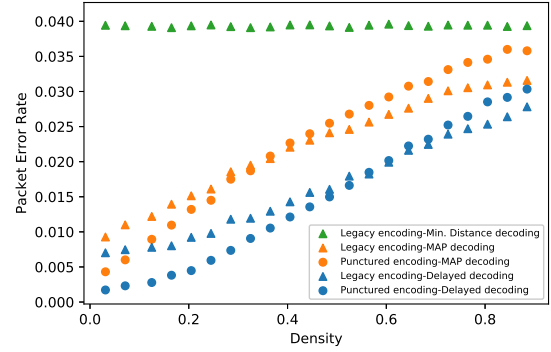}
  \caption{Packet error probability as a function of the density of the transition matrix $(p_b=0.05)$. Delay $d=1$ in all cases of delayed decoding}
  \label{Fig:Error_vs_density}
  \end{center}
\end{figure}

Fig.~\ref{Fig:Error_vs_density} shows the effect of the sparsity of transition matrix on packet error probability once steady state is reached. Unsurprisingly, minimum distance decoding does not depend on sparsity of the transition matrix, while other decoding schemes can exploit their knowledge of the Markovian source to have much better results for sparser matrices, i.e., for less dynamic sources whose state is highly correlated in time. Punctured encoding is designed to exploit the knowledge about the Markov source; hence, it performs better than legacy encoding for sparser transition matrices, but becomes inefficient if the transition matrix is very dynamic. In realistic \gls{iot} scenarios, we expect transition matrices to be relatively sparse, as sensors often measure processes such as temperature or air pollution, which are highly correlated in time. In this case, compressed encoding schemes performed worse than either legacy or punctured encoding, so they were not included in the plot.

\begin{figure}[t!]
\begin{center}
  \includegraphics[width=0.9\columnwidth]{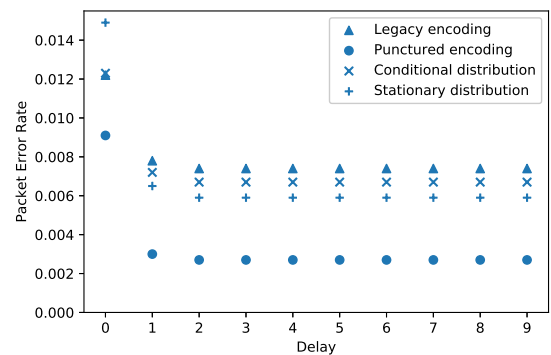}
  \caption{Packet error probability as a function of the delay $d$ for combined encoding  $(p_b=0.05$ and $T$ is a sparse matrix with density 0.125$)$}
  \label{Fig:Error_vs_delay}
  \end{center}
\end{figure}

As Fig.~\ref{Fig:Error_vs_delay} shows, the benefits of delayed decoding are significant even with a delay of just 1 packet. Further improvements from increasing the delay to 2 or more packets become negligible, as the state of the source is Markovian.

\begin{figure}[t!]
\begin{center}
  \includegraphics[trim=0 0 0 0.4cm,width=0.9\columnwidth]{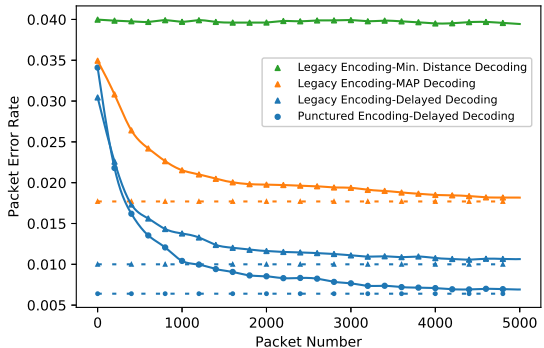}
  \caption{Packet error probability as a function of the number of packets received for various schemes. The dashed line represent the packet error probability if the receiver perfectly knows the transition matrix. $(T$ is a sparse matrix with density 0.25$)$. Delay $d=1$ in all cases of delayed decoding}
  \label{Fig:Error_vs_time}
  \end{center}
\end{figure}

Fig.~\ref{Fig:Error_vs_time} shows performance before the system reaches steady state, i.e., how packet error probabilities for different schemes vary over time if the system starts with no knowledge of the source parameters. The learning parameter $\alpha$ has been set to $0.1$, which showed the best results. The dashed lines represent the perfect-knowledge packet error probabilities, which assume that the receiver has perfect knowledge of the transition matrix. Source compression is not an option here, as we first have to learn the statistics of the source to use those encoding schemes. The figure shows that the system converges to a packet error rate very close to the perfect-knowledge performance after 2000 packets, and that punctured encoding has a better performance than the legacy scheme during all phases of the initial transition to steady state.

\begin{figure}[t!]
\begin{center}
  \includegraphics[trim=0 0 0 0.4cm, width=0.9\columnwidth]{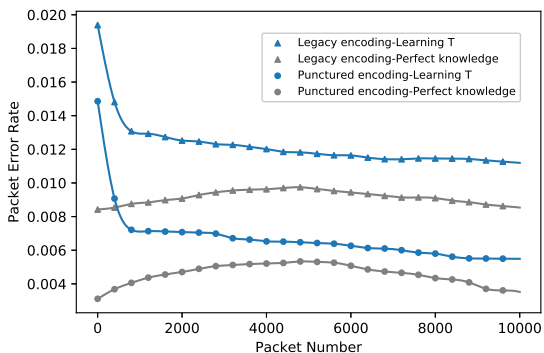}
  \caption{Packet error probability as a function of the number of packets received for a dynamic source with delayed decoding. The grey plots show the perfect knowledge performance, while the blue plots represent when receiver learns the transition matrix with time. $T_1$ and $T_2$ are sparse matrices with density $0.125$ and delay $d=1$}
  \label{Fig:Dynamic}
 \end{center}
\end{figure}

Finally, we simulate a dynamic source, whose underlying Markov chain changes over time: two sparse matrices $(T_1$ and $T_2)$ are generated randomly and the transition matrix of the source at any moment is a convex combination of $T_1$ and $T_2$.
\begin{equation}
    T=\left(1-\frac{t}{t_{total}}\right)T_1+\frac{t}{t_{total}}T_2
\end{equation}
where $t_{total}$ is the total number of packets sent ($t_{total}=10000$ in Fig.~\ref{Fig:Dynamic}) and $t$ is the packet number of each packet. This process is repeated for 500 sequences and the averaged results are shown in Fig.~\ref{Fig:Dynamic}. As the figure shows, our scheme can deal with a time-varying transition matrix, achieving most of the benefits of perfect knowledge of the transition matrix. In this kind of system, source compression is not an option, as the source would have to periodically renew its encoding of the state and transmit the map to the receiver to allow decoding. The higher packet error rate in the middle is due to the lower sparsity of the transition matrix caused by the linear combination of the two matrices.

\section{Conclusion and future work}\label{sec:concl}

In this paper, we have presented a puncturing-based encoding and decoding scheme that improves the transmission of short packets from time-correlated sources, modeled as \glspl{hmm}. Our scheme significantly outperforms compression-based approaches, especially for more stable sources.

There are several avenues of future work that we plan to follow: the extension of the scheme to \gls{harq} schemes, with partial retransmission of the data in case of failure, is an interesting possibility, as is its natural extension to multiple sources of information. The state of each source would be exploited to improve the source identification, reducing the need for long packet headers, and the correlation between adjacent sources could be exploited to improve reliability even further, adding a spatial component as well as the temporal one we analyzed in this work.

\section*{Acknowledgment}
 This work has been in part supported by the Danish Council for Independent Research, Grant Nr. 8022-00284B SEMIOTIC.

\bibliographystyle{IEEEtran}
\bibliography{IEEEabrv,references}

\begin{thebibliography}{10}
\providecommand{\url}[1]{#1}
\csname url@samestyle\endcsname
\providecommand{\newblock}{\relax}
\providecommand{\bibinfo}[2]{#2}
\providecommand{\BIBentrySTDinterwordspacing}{\spaceskip=0pt\relax}
\providecommand{\BIBentryALTinterwordstretchfactor}{4}
\providecommand{\BIBentryALTinterwordspacing}{\spaceskip=\fontdimen2\font plus
\BIBentryALTinterwordstretchfactor\fontdimen3\font minus
  \fontdimen4\font\relax}
\providecommand{\BIBforeignlanguage}[2]{{%
\expandafter\ifx\csname l@#1\endcsname\relax
\typeout{** WARNING: IEEEtran.bst: No hyphenation pattern has been}%
\typeout{** loaded for the language `#1'. Using the pattern for}%
\typeout{** the default language instead.}%
\else
\language=\csname l@#1\endcsname
\fi
#2}}
\providecommand{\BIBdecl}{\relax}
\BIBdecl

\bibitem{forecast2019cisco}
Cisco, ``Visual {Networking Index}: global mobile data traffic forecast update,
  2017--2022,'' Tech. Rep., 2017.

\bibitem{shi2020low}
Y.~Shi, J.~Dong, and J.~Zhang, \emph{Low-overhead Communications in {IoT}
  Networks}.\hskip 1em plus 0.5em minus 0.4em\relax Springer, 2020.

\bibitem{wu2020pilot}
J.~Wu, W.~Kim, and B.~Shim, ``Pilot-less one-shot sparse coding for short
  packet-based machine-type communications,'' \emph{IEEE Transactions on
  Vehicular Technology}, May 2020.

\bibitem{rastogi2020iot}
K.~Rastogi, A.~Barthwal, D.~Lohani, and D.~Acharya, ``An {IoT}-based discrete
  time {Markov} chain model for analysis and prediction of indoor air quality
  index,'' in \emph{Sensors Applications Symposium (SAS)}.\hskip 1em plus 0.5em
  minus 0.4em\relax IEEE, 2020, pp. 1--6.

\bibitem{shirbhate2018time}
I.~M. Shirbhate and S.~S. Barve, ``Time-series energy prediction using {Hidden
  Markov Model} for smart solar system,'' in \emph{2018 3rd International
  Conference on Communication and Electronics Systems (ICCES)}.\hskip 1em plus
  0.5em minus 0.4em\relax IEEE, 2018, pp. 1123--1127.

\bibitem{rabiner1989tutorial}
L.~R. {Rabiner}, ``A tutorial on hidden {Markov} models and selected
  applications in speech recognition,'' \emph{Proceedings of the IEEE},
  vol.~77, no.~2, pp. 257--286, 1989.

\bibitem{stojkoska2017data}
B.~R. Stojkoska and Z.~Nikolovski, ``Data compression for energy efficient iot
  solutions,'' in \emph{25th Telecommunication Forum (TELFOR)}.\hskip 1em plus
  0.5em minus 0.4em\relax IEEE, Nov. 2017, pp. 1--4.

\bibitem{biason2017ec}
A.~Biason, C.~Pielli, M.~Rossi, A.~Zanella, D.~Zordan, M.~Kelly, and M.~Zorzi,
  ``{EC-CENTRIC}: An energy-and context-centric perspective on {IoT} systems
  and protocol design,'' \emph{IEEE Access}, vol.~5, pp. 6894--6908, Apr. 2017.

\bibitem{islam2018information}
N.~S. Islam and W.~Henkel, ``Information forwarding in {LDPC} decoding for
  {Markov} sources,'' in \emph{10th International Symposium on Turbo Codes \&
  Iterative Information Processing (ISTC)}.\hskip 1em plus 0.5em minus
  0.4em\relax IEEE, Dec. 2018, pp. 1--5.

\bibitem{thobaben2005low}
R.~Thobaben and J.~Kliewer, ``Low-complexity iterative joint source-channel
  decoding for variable-length encoded {Markov} sources,'' \emph{IEEE
  Transactions on Communications}, vol.~53, no.~12, pp. 2054--2064, Dec. 2005.

\bibitem{majumder2012joint}
S.~Majumder and S.~Verma, ``Joint source-channel decoding of {IRA} code for
  hidden {Markov} source,'' in \emph{1st International Conference on Recent
  Advances in Information Technology (RAIT)}.\hskip 1em plus 0.5em minus
  0.4em\relax IEEE, Mar. 2012, pp. 220--223.

\bibitem{bose1960class}
R.~C. Bose and D.~K. Ray-Chaudhuri, ``On a class of error correcting binary
  group codes,'' \emph{Information and control}, vol.~3, no.~1, pp. 68--79,
  Mar. 1960.

\bibitem{zhao2006turbo}
Y.~Zhao and J.~Garcia-Frias, ``Turbo compression/joint source--channel coding
  of correlated binary sources with hidden markov correlation,'' \emph{Signal
  Processing}, vol.~86, no.~11, pp. 3115--3122, 2006.

\bibitem{garcia2003ldpc}
J.~Garcia-Frias and W.~Zhong, ``{LDPC} codes for compression of multi-terminal
  sources with hidden {Markov} correlation,'' \emph{IEEE Communications
  Letters}, vol.~7, no.~3, pp. 115--117, Mar. 2003.

\end{thebibliography}
\end{document}